# Giant ultra-broadband photoconductivity in twisted graphene heterostructures

H. Agarwal[1]*, K. Nowakowski[1]*, A. Forrer[2], A. Principi[6], R. Bertini[1], S. Batlle-Porro[1], A. Reserbat-Plantey[1,3], P. Prasad[1], L. Vistoli[1], K. Watanabe[4], T.Taniguchi[,5], A. Bachtold[1,7], G. Scalari[2], R. Krishna Kumar[1]+, F. H. L. Koppens[1,7]+.

1 ICFO - Institut de Ciencies Fotoniques, The Barcelona Institute of Science and Technology, 08860 Castelldefels (Barcelona), Spain.

2. Quantum Optoelectronics Group, Institute of Quantum Electronics, ETH Zürich, 8093 Zürich, Switzerland

3. Université Côte d'Azur, CNRS, CRHEA, 06560 Valbonne, France

4. Research Center for Functional Materials, National Institute for Materials Science, Tsukuba, Japan

5. International Center for Materials Nanoarchitectonics, National Institute for Materials Science, Tsukuba, Japan

6. School of Physics and Astronomy, University of Manchester, Manchester M13 9PL, United Kingdom

7. ICREA—Institució Catalana de Recerca i Estudis Avançats, Barcelona, 08010, Spain

*These authors contributed equally

+correspondence: roshan.krishnakumar@icfo.eu, frank.koppens@icfo.eu

**The requirements for broadband photodetection are becoming exceedingly demanding in hyperspectral imaging. Whilst intrinsic photoconductor arrays based on mercury cadmium telluride represent the most sensitive and suitable technology, their optical spectrum imposes a narrow spectral range with a sharp absorption edge that cuts their operation to < 25 µm. Here, we demonstrate a giant ultra-broadband photoconductivity in twisted double bilayer graphene heterostructures spanning a spectral range of 2-100 µm with internal quantum efficiencies ~ 40 % at speeds of 100 kHz. The giant response originates from unique properties of twist-decoupled heterostructures including pristine, crystal field induced terahertz band gaps, parallel photoactive channels, and strong photoconductivity enhancements caused by interlayer screening of electronic interactions by respective layers acting as sub-atomic spaced proximity screening gates. Our work demonstrates a rare instance of an intrinsic infrared-terahertz photoconductor that is complementary metal-oxide-semiconductor compatible and array integratable, and introduces twist-decoupled graphene heterostructures as a viable route for engineering gapped graphene photodetectors with 3D scalability.**

Broadband operation is the property of photodetectors that not only expands their functionality but is an essential requirement for spectroscopic applications. Moreover, there is a growing need for photodetectors operating over multiple spectral regimes, for example, in observational astronomy[1]. Intrinsic photoconductors offer the highest responsivity and fast operation but suffer from a sharp absorption edge so that applications require the use of multiple devices optimized for different wavelengths, significantly lowering the pixel density of photodetector arrays[2]. Bolometers on the other hand offer broadband response extending to > 100 µm but suffer from low dynamic range and dark current that limits the noise-equivalent-power, and require challenging array integration processes. Hence, a high-quality narrow-gapped photoconductor could offer a route towards high responsive, fast, broadband photodetectors bridging the challenging detection range between 10 and 100 µm. Although graphene offers great opportunity for broadband operation[3], its gapless spectrum inhibits any sizeable photoconductive response. It instead relies on junction phenomena[4] with small optically active areas or engineering gapped[5,6]/resistive[7] behaviour in graphene which compromise its desirable electronic quality and requires complicated device architectures. Furthermore, its inherent 2D nature severely limits extrinsic light absorption, making it impossible to compete with conventional 3D photoconductors whose device thickness is optimized to around 10 µm[8].

In this work, we show that large angle-twisted double bilayer graphene (TDBG) (Fig. 1a) overcomes such restrictions and can be used to engineer high quality broadband photoconductors spanning multiple spectral ranges. In contrast to conventional semiconductors that typically find maximum response for excitation energies at the band gap, graphene bilayers are unique in that they exhibit two absorption edges corresponding to the transitions between interlayer interaction induced band splitting[9] at ∼ 400 meV (labelled *A1* Fig. 1b) and, if inversion symmetry is broken, a semiconducting bandgap at the K point[10] (labelled *A2* Fig. 1b). In large-angle TDBG, twist-decoupling of the graphene bilayers results in a pristine narrow bandgap (~ 10 meV) caused by crystal fields[11] intrinsic to heterostructure, pushing A2 into the terahertz (THz) regime. The unique combination of two absorption edges allows for an ultra-broadband photoconductive response from 2-100 µm unparalleled by any other intrinsic photoconductor[8]. Furthermore, the crystal field gap circumvents the need for dual-gated device architectures[5] and provides a route towards 3D scaling of graphene photodetectors with 2D properties. We demonstrate this possibility by showing individual bilayers within the heterostructure behave as parallel photoconductive channels which increases the optical volume and extrinsic absorption whilst preserving, and even enhancing, the favourable properties of 2D crystals. Specifically, we find a giant photoconductivity enhancement of bilayer graphene (BLG) sheets placed in the twist-decoupled environment caused by interlayer screening of electron-hole collisions. The twist-decoupling of individual layers combined with the crystal field gaps has potential for vertical scaling of gapped graphene photoconductors with extrinsic device performances surpassing 3D photoconductors.

Our devices were fabricated using standard fabrication techniques[12,13] (see methods). The studied heterostructures consisted of large-angle (> 15°) TDBG encapsulated with hexagonal boron-nitride (hBN) and placed on top of either graphite gates or silicon gates. Some of the devices had top gates to characterize the system in accordance with previous works[11] (see Supplementary Section 1) but were not essential for device operation. The blue trace in Fig. 1c plots the resistivity ($\rho_{xx}$) as a function of carrier density (*n*) measured in one of our TDBG devices, showing a peaked response at the charge neutrality point (CNP), *n* = 0. This value is rather large compared to a typical zero density of states peak observed in graphene devices and indicates the presence of a gap. Temperature dependent measurements reveal the expected activation behaviour from which gaps of ∼ 10 meV are extracted in agreement with previous works[11] (Supplementary Section 1). The presence and high quality of the gapped state was found robust between different devices (see Supplementary Section 2).

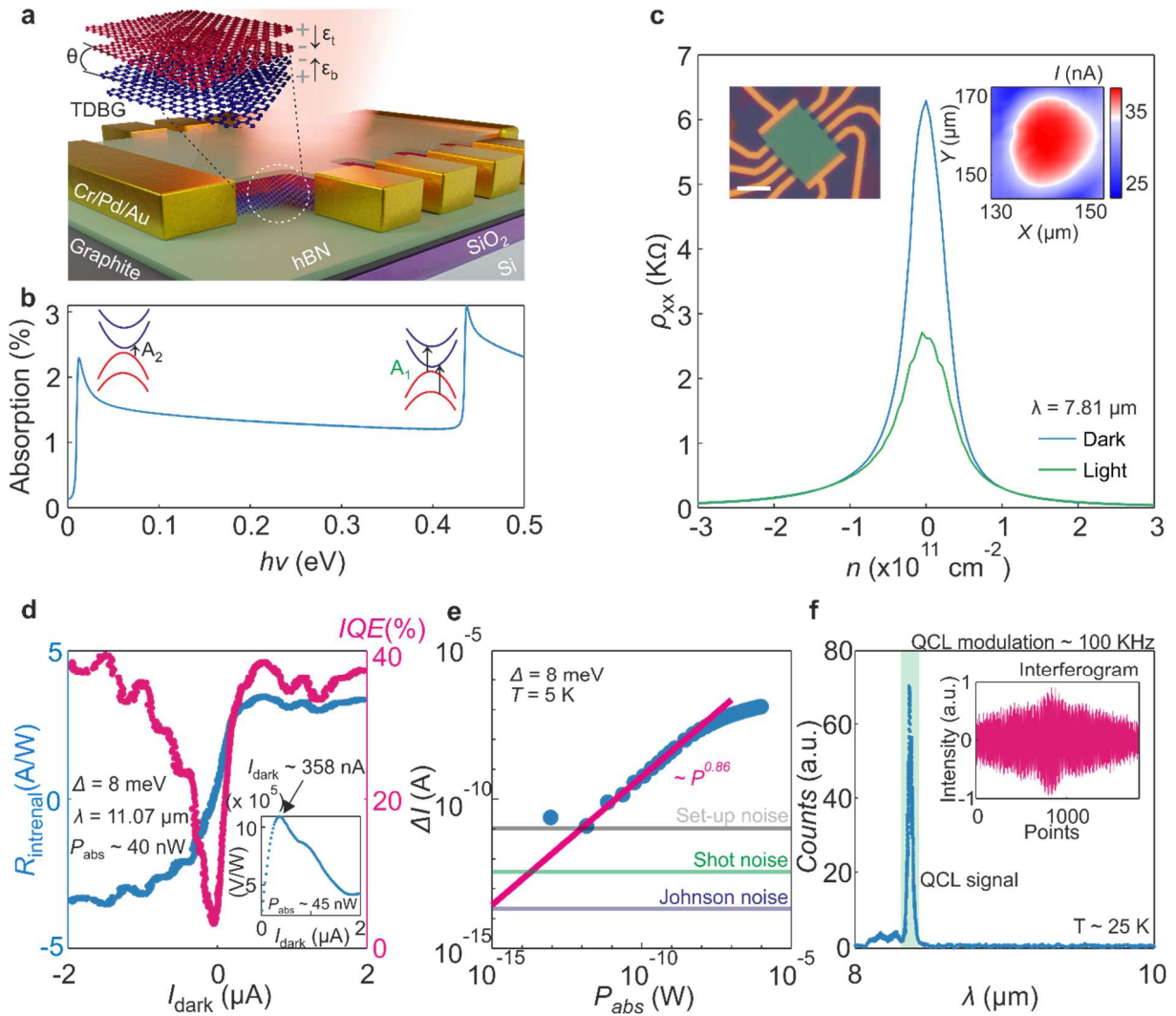

**Figure. 1 Giant photoconductive response in large angle TDBG. a,** Device schematic illustrating the studied heterostructures consisting of hBN (green), top BLG (Red) and bottom BLG (Blue) crystal layers. The arrows indicate the direction of the displacement fields that gap the bilayers due to crystal fields intrinsic to the heterostructure. **b,** Calculated absorption spectrum as a function of photon energy for pristine bilayer graphene (see Supplementary Section 3). Inset cartoons show band structure of TDBG. The green and black arrows highlight the resonant transitions caused by interlayer interaction band splitting (A1) and crystal field induced band gap (A2) respectively. **c,** Resistivity ($\rho_{xx}$) as a function of gate-induced carrier density ($n$) measured in our TDBG heterostructure with (green) and without (blue) photoexcitation at 7.81 μm (absorbed power ~ 30 μW, and absorbed power density ~ 0.2 W/μm$^2$). Inset top left: Optical image of a TDBG device, with scale bar of 10 μm. Top right inset: spatial map of the photoresponse plots photoconductivity ($\Delta V$) as a function of laser spot spatial position $x$ and $y$. **d,** Responsivity and quantum efficiency plotted as a function of DC current bias. **e,** Two probe photocurrent measured as a function of absorbed power at an optimal bias of 17 mV. The solid magenta line represents a power fit done in the linear response region, and the horizontal lines represent different noise levels. **f,** Fourier transform of the Interferogram (inset) measured whilst

excitation with a quantum cascade laser (QCL) in our TDBG device. The photo voltage measurement was locked to the frequency of the QCL which was operated at 100 KHz, demonstrating a lower bound estimate of the device speed.

**Ultra-broadband photoconductivity in TDBG**

To characterize the photoresponse of the device, we first performed scanning photoconductivity measurements at 7.81 µm excitation. The green curve in Fig. 1c plots $\rho_{xx}$ as a function of $n$ under continuous illumination (see methods). It shows a strong photoconductive response in which the resistivity at the CNP decreases dramatically upon photoexcitation. Spatially resolved measurements (top right inset of Fig. 1c) shows that the photoresponse is concentrated in the bulk of the sample, demonstrating that the entire device area is photoactive. By optimizing the bias, we found a linear response for a large dynamic range (Fig. 1e), intrinsic current/voltage responsivities of $4AW^{-1}/10^6 \, VW^{-1}$, internal quantum efficiency of 40 % (Fig. 1d), set-up limited speed measured at 100 kHz (Fig. 1f), a linear dynamic range of 80 dB (Fig. 1e) and setup-limited noise equivalent powers of 1 pW/$\sqrt{Hz}$, while the intrinsic noise equivalent power is estimated at 66 fW/$\sqrt{Hz}$. These are competitive detector parameters compared with commercial photoconductor technologies operating in the infrared range (IR) (see Supplementary Section 3 for details on detector parameter calculations). Similarly high response could be observed in other studied TDBG devices (see Supplementary Section 2).

Because of the unique optical conductivity of graphene bilayers, we expect a broadband operation extending into the far infrared regime until the second absorption edge (A2). To study the spectral response, we employed Fourier transform infrared photocurrent spectroscopy (FTIR)[14] with a broadband thermal source (see methods). Figure. 2a plots the measured photocurrent spectrum of our device covering the range 2- 16 µm (20-150 THz); the magenta data points plot the photocurrent $I_{PC}$ as a function of frequency ($\nu$) normalized by the power of the source. The response in the dark is plotted in aqua to highlight the noise level. Notably, we find a broadband response covering the entire spectral range with features reflecting the optical conductivity of our TDBG heterostructure. The main spectral peak from 2-16 µm reflects the absorption between parallel bands[9] (A1) and notable dips around 32 and 40 THz are attributed to enhanced absorption in the silicon and hBN optical phonon bands respectively. Even for the lowest frequencies the response remains finite but the spectral range is cut-off by the ZnSe windows on the cryostat.

To explore the long wavelength regime, we performed measurements using THz sources. As a first step, we characterised the THz photoconductivity using a monochromatic source at 2.5 THz (10 meV). Like for the IR measurements, we found a peaked response at the CNP and a linear power dependence indicative of a photoconductive effect (see Supplementary Section 4). Following, we extended the FTIR photocurrent measurements to THz frequencies using a THz QCL as a source[15]. Figure. 2b plots the spectral response of our device in magenta solid lines, compared with the spectral response of a commercial detector (solid aqua lines). The two peaked structures seen in the commercial detector at 2.5 THz and 3.4 THz represent the main emission lines of the QCL. Strikingly, our devices follow well the spectral response of the peak centred on 3.4 THz demonstrating its broadband operation still at THz frequencies. We also note that the photoresponse was measured by locking to the repetition rate of the QCL, demonstrating a lower bound operating speed of 10 kHz even at THz frequencies. However, our device showed no responsivity around the second peak centred at 2.3 THz. These frequencies correspond roughly to the band gaps measured in our quantum transport studies and suggests that we reach the absorption edge of the crystal field gap in TDBG, marking the cut-off frequency of our devices.

The above measurements demonstrate the exceptional broadband nature of our TDBG photoconductors spanning IR to THz regimes. Knowing the absorbed power (see Supplementary Section 3) we can determine the frequency dependent internal quantum efficiency (IQE). Figure. 2c plots the IQE (left axis) as a function of wavelength for our TDBG device (cyan circles) and a graphene photoconductor (cyan square) measured in ref (16). IQE calculations for wavelengths > 18 µm was not possible because estimations of the absorbed power became non-trivial; the solid cyan line plots a lower bound estimation based on the absorption spectrum of our TDBG heterostructure and measured terahertz response (Fig. 2b). To benchmark the spectral response, we plot the quantum efficiency of different photoconductor technologies on the right axis in magenta; detective quantum efficiencies are plotted only to compare the spectral response. Our TDBG devices show a unique spectral shape in the quantum efficiency. First, the IQE is remarkably high close to A1 even exceeding unity. Whereas 100% IQE is typical for a high-quality photoconductor, exceeding unity suggests the presence of gain. Given that the photo-excitation energy is significantly larger than the band gap (A2) carrier multiplication may play a role. Indeed, the IQE shows a linear dependence with photon energy (Inset of Fig. 2c) hinting at carrier multiplication present in our devices[17]. Second, although the IQE drops away from A1, it remains large and finite > 30 % over the entire spectral range. This is in contrast to typical photodetectors in the IR spectral range that exhibit a sharp cut-off to zero QE below the gap (Fig. 2c).

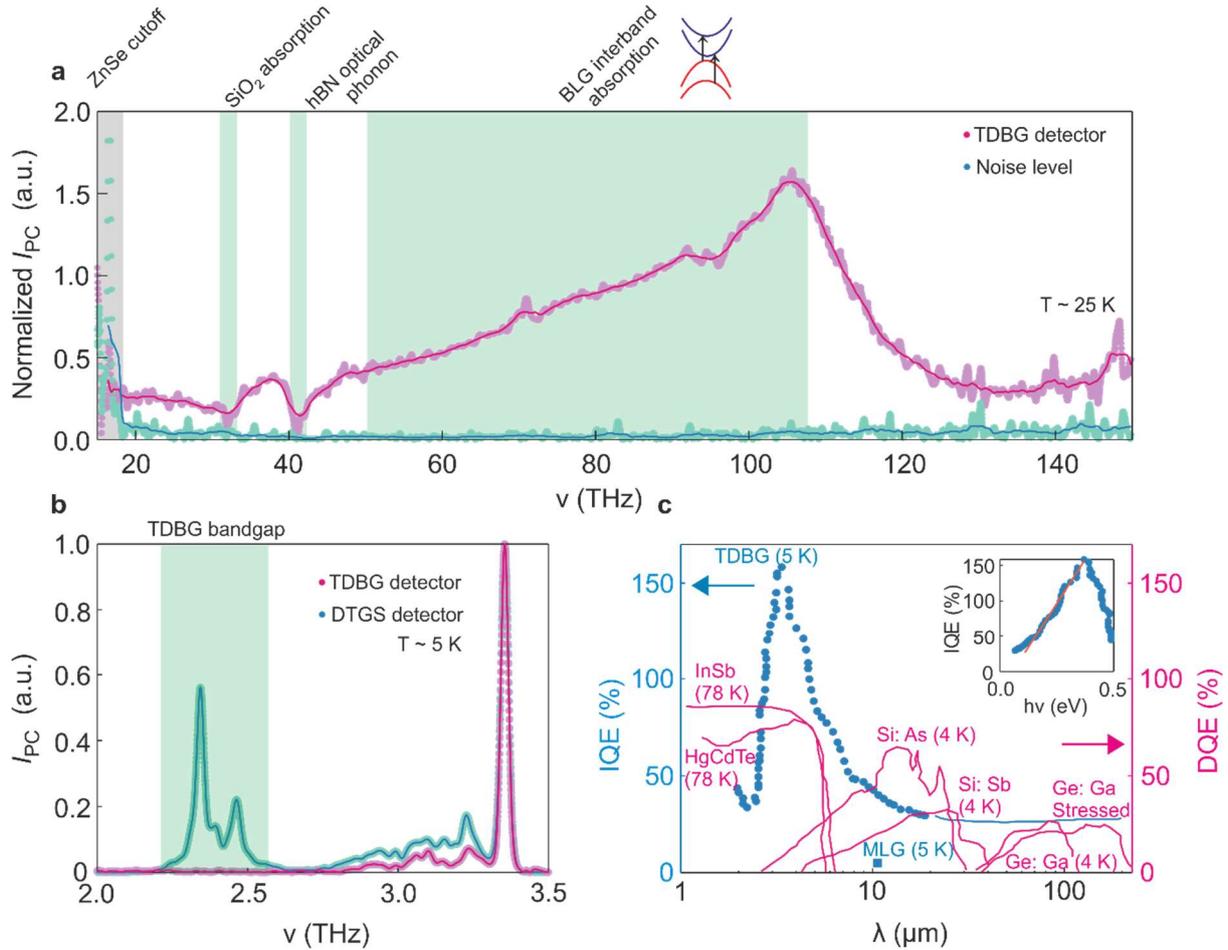

**Figure. 2 ultra-broadband spectral response from infrared to terahertz wavelengths in TDBG photodetector**. **a,** FTIR photocurrent spectrum measured at 25 K, data is normalized by the spectrum measured with the deuterated triglycine sulfate (DTGS) detector. **b,** FTIR photocurrent spectrum measured at 5 K using a THz QCL modulated at 10 kHz. Aqua and magenta lines plot the response from the DTGS and our TDBG detector respectively (Maxima of both spectrums are normalized to unity). Light green shaded region indicates energies below the second absorption edge A2 (10 meV) corresponding to the crystal field induced bandgap in TDBG. **c,** IQE as a function of wavelength plotted on the left axis in cyan circles represent our TDBG device, and the cyan square represents previous work[16]. On the right axis, the detective quantum efficiency (DQE) is plotted for various photoconductor technologies in magenta[18]. Top right inset: Zoom in of IQE as a function of photon energy (up to 2.5 μm).

**Giant photoconductivity enhancements in twisted graphene heterostructures**

Although small, the crystal field induced gap (~ 10 meV) in our devices inflicts a sizeable response with extrinsic voltage responsivities one order of magnitude larger than previous works on dual-gated bilayer graphene[5] that operated with 5 x larger gap sizes (50 meV) and dark resistivities (100 kΩ). Whilst we do expect higher response in TDBG due its larger optical volume, such an improvement cannot be explained by absorption enhancements alone. To understand the origin, we performed control experiments in hBN encapsulated dual-gated BLG heterostructures with gaps tuned to ∼ 10 meV (same as our TDBG device, see Supplementary Section 5). Fig. 3 plots power dependent measurements of the photoconductivity $\Delta\sigma_{xx} = \sigma_{Light} - \sigma_{Dark}$ for one of our TDBG and BLG devices. Strikingly, it shows $\Delta\sigma_{xx}$ significantly larger in our TDBG device for the same power range. Indeed, a factor of two enhancement can be expected due to the larger optical volume – our TDBG consists of two photoactive bilayers and so absorbs twice the amount of light. However, we find doubling the number of layers causes a 4-fold increase. This suggests individual bilayers themselves experience a 100 % photoconductivity enhancement in TDBG. This is quite surprising considering the layers are electronically decoupled[11] (see Supplementary Section 1), yet their opto-electronic properties are somehow enhanced in the twisted environment.

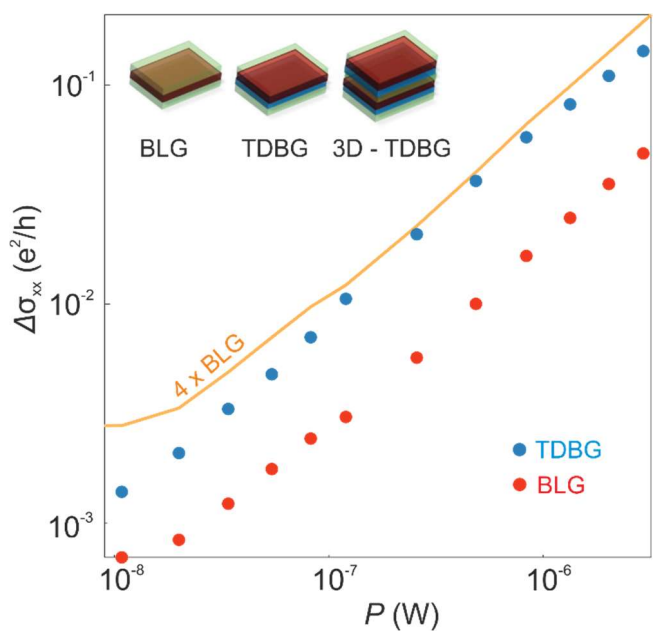

**Figure. 3 Giant photoconductivity enhancement in TDBG.** Photoconductivity ($\Delta\sigma_{xx}$) as a function of illuminated power (*P*) measured in one of our TDBG devices (blue circles) compared with a control BLG sample (red circles). Both devices had gaps at the gamma point of ∼ 10 meV. Measurements performed at 5 K under illumination with light of wavelength 11.07 μm. Top left inset: Schematics of BLG, TDBG, and envisioned 3D-TDBG heterostructures. While going to 3D structures, it is important to have insulating layer between two TDBG to maintain the crystal field.

To understand the photoconductivity enhancement, we first proceed to understand the mechanism. There are two possible mechanisms that can describe the photoresponse, either the photoconductive effect or the bolometric effect. Decoupling the two mechanisms in gapped systems is non-trivial because both effects lead to experimentally similar consequences – free carriers in the conduction and valence band. However, the spectral response that mimics the optical conductivity provides strong evidence that the mechanism originates from photoconductivity; photon detectors are dependent on the number of photons absorbed whereas bolometers depend only on the total energy density and exhibit a flat spectral response. The two mechanisms are also distinguished in the IV characteristics of the photoresponse by different Fermi-Dirac distributions of electrons and holes. In the case of a bolometer, the carriers should be described by one Fermi-Dirac distribution (Inset Fig. 4a), whereas in a photoconductor, photoexcited carriers lead to separate electron and hole distributions described by two *quasi*-Fermi levels (Inset Fig. 4b). A simple model (see Supplementary Section 6) shows that the two distributions lead to rather different behaviour in the IV characteristics of gapped systems (Fig. 4a, b). In the case of a bolometer, temperature acts to smear the IV curve in the region where the bias voltage approaches the gap (Fig. 4a). However, in photoconductivity, the system attains a finite conductance even close to zero bias. Hence, the key differences can be seen close to zero bias where in a bolometer this value remains small for a certain temperature range and only becomes finite when $T$ smears the Fermi-Dirac distribution across the gap, whilst in the photoconductor the conductivity increases immediately due to an increase in the carrier density by photoexcitation. The differences can be seen clearer in plots of the derivative (Inset in Fig. 4a, b). To compare, we plot the IV characteristics measured in our TDBG device for different powers in Fig. 4c. It clearer resembles that expected from a photoconductor with two separate electron and hole Fermi-Dirac distributions.

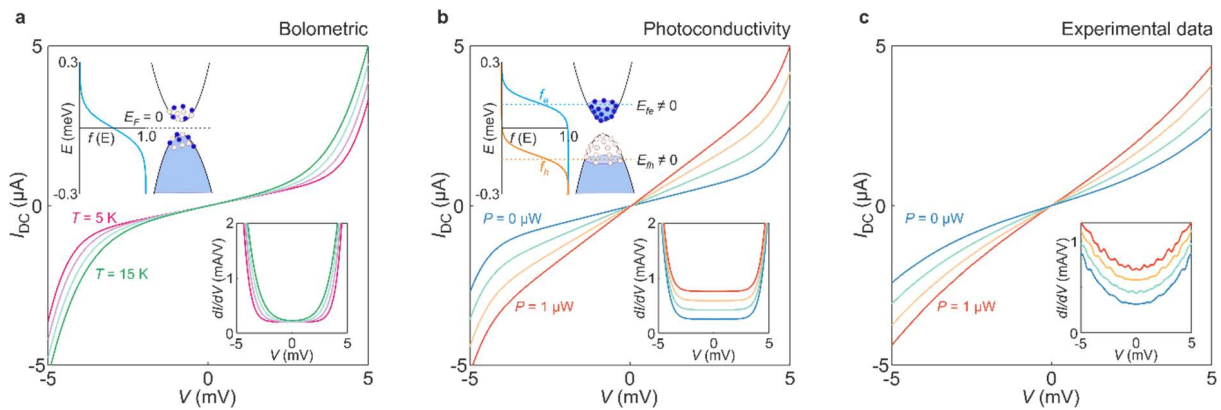

**Figure. 4 Photoconductivity Mechanism: a**, Simulated IV curves for a bolometric effect in our devices. The data describes the effects of temperature increase by laser heating on the IV characteristics of a gapped system. The finite slope around zero bias describes the dark current present in our devices. The input parameters are ($T$ = 5 K, $R_{dark\ resistance}$ = 10 kΩ, $\Delta$ = 10 meV). Top left inset: illustration of a single Fermi-Dirac distribution. Bottom right inset: derivatives of the data in main panel ($dI/dV$). **b,** simulated IV curves for a photoconductive effect in which light illumination pumps carriers into valence and conduction bands with separate Fermi-Dirac distributions and quasi-Fermi levels. Top left inset: illustration of carrier distribution in the case of a photoconductor represented by two Fermi-Dirac distrbutions. Bottom right inset: derivatives of the data in main panel ($dI/dV$). **c,** experimental data plotting IV curves measured under different laser illumination intensities with $\lambda$ = 11.07 μm. Bottom right inset: derivative of the data in main panel. IV curves plotted for increasing powers corresponding to a photoconductive effect. Top left inset illustrates the probability density function of separate electron and hole Fermi Dirac distributions.

According to a photoconductive model, there are only two material parameters that can influence $\Delta \sigma_{xx}$ which are the mobility ($\mu$) or the recombination lifetime ($\tau$), according to

$$\Delta \sigma_{xx} = \Delta n e \mu$$

$$\Delta n \sim \tau,$$

where $\Delta n$ is the photoexcited carrier density governed by $\tau$. Although $\tau$ can vary with power[19], we focus here on the linear response regime with constant $\tau$. Further, it is unlikely that $\tau$ is enhanced in TDBG, since an increase in the number of layers generally increases the number of recombination channels. Therefore, μ must be somehow enhanced. To understand if any differences in the mobility may be present between BLG in the twisted heterostructure and stand-alone BLG, we performed DC transport measurements focusing at the CNP ($n = 0$) where photoconductivity was measured. Figure 5 plots the conductivity ($\sigma_{xx}$) as a function of temperature ($T$) measured at the CNP for all our studied devices plotted with different symbols in magenta and blue for BLG and TDBG respectively; gaps in all devices ∼ 10 meV. The transparent magenta line plots twice the BLG conductivity to highlight any conductivity enhancement. Notably, above 30 K the TDBG conductivity starts to deviate from the 2 x BLG line indicating a higher mobility in individual layers. At low $T$ it is hard to compare different devices due to large variations in sub-gap conductivity[20]. However, at high $T$ the conductivity of BLG has been shown to be $T$ independent and universal governed by electron-hole collisions with the scattering rate approaching Planckian dissipation[21,22]. Indeed, our BLG devices show similar universal behaviour with the conductivity saturation to around 21 $e^2$/h. Strikingly, in our TDBG devices we found similar saturating behaviour but to values around 3 x BLG. At such high temperatures, we cannot attribute mobility enhancements due to extrinsic effects and the results show that e-h collisions are somehow supressed in our TDBG heterostructure.

One possibility to control Coulomb interactions is through screening effects via local proximity screening gates[23]. Such effects are hard to observe in BLG due to its parabolic spectrum requiring separation distances of ∼ 7 Å [23], and thus it is unlikely that local graphite gates could have such an effect because they are spaced ∼ 20 nm away. This is consistent with the fact that photoconductivity in our BLG devices on Si gates and graphite gates showed rather similar behaviour (See Supplementary Section 7). However, because the bilayers are decoupled within the heterostructure, each bilayer can be seen as a proximity screening gate for the other. Recent work has shown that decoupled bilayers can be modelled as a parallel plate capacitor separated by an air gap with an effective dielectric constant[11] (Inset of Fig. 5), and there have been some reports on the effects of screening in graphene double layers[24,25]. Making an estimate for the correction to the e-e scattering length using the formula derived in ref. 19, taking the Thomas-Fermi screening length in BLG and an effective interlayer distance thickness of 2.6 Å [26], we find enhancement factors close to what is observed experimentally ∼ 1.5 - 2 (see Supplementary Section 8). With this in mind, we made a full calculation of the DC conductivity in twist-decoupled bilayer heterostructures (see Supplementary Section 9) and derive an analytical formula for the conductivity. Remarkably, we find that doubling the number of layers within the heterostructure indeed should cause a fourfold increase in the conductivity. This is because doubling the number of layers not only doubles the carrier density $n$, but the additional carriers screen interactions and increase the mobility by a factor of two, causing a four-fold enhancement of the conductivity when doubling the number of bilayers, as observed in our photoconductivity experiments (Fig. 3). This suggests that the photoconductivity enhancements observed in TDBG originates from interlayer screening of e-h collisions.

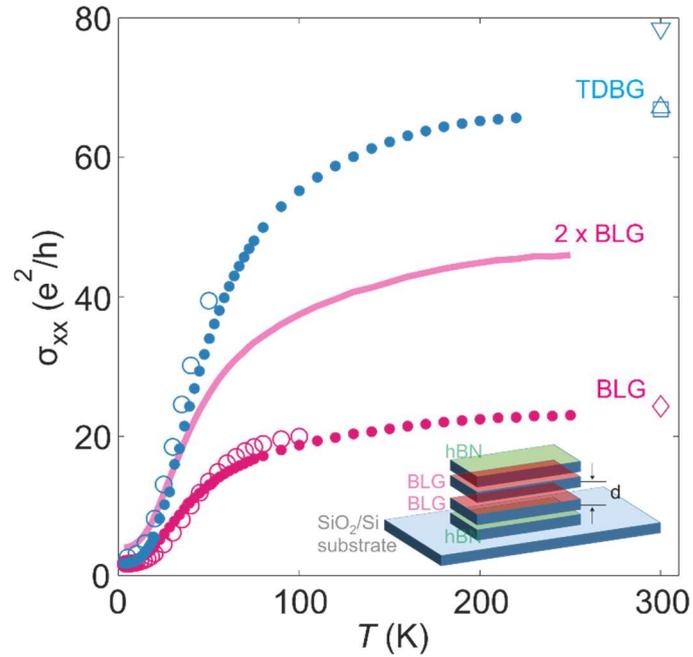

**Figure. 5 Interlayer screening of e-h collisions in TDBG**. DC conductivity ($\sigma_{xx}$) measured as a function of temperature ($T$) at the neutrality point ($n$ = 0) in all our TDBG and BLG samples with gaps of ∼ 10 meV. The blue and magenta colour coding corresponds to TDBG and BLG devices respectively and the different shapes correspond to different devices measured. The solid transparent magenta line traces twice the measured BLG conductivity. Inset shows a cartoon sketch of the TDBG device encapsulated in top and bottom hBN. Here, top and bottom BLG are separated by atomic vacuum space (d).

### Discussion

The above measurements demonstrate the suitability of large-angle TDBG for ultra-broadband photo detection with large quantum efficiencies and fast modulation extending to terahertz frequencies. Whilst we measure a lower bound speed of 100 kHz, the speed is limited only by the lifetime of carriers $\tau$ and therefore could operate at megahertz and even gigahertz frequencies. Although the extrinsic device absorption is still small compared to commercial photoconductors, Fig. 3 demonstrates the suitability of twist de-coupled graphene heterostructures for increasing the optical volume of devices whilst preserving favourable 2D properties including the gapped state. There are several ways to reliably gap graphene but it becomes exponentially more difficult to preserve the gap when increasing the number of layers. Superlattice requires alternating layers of angle specific aligned crystal layers[6] whereas dual-gated architectures[27] require complex electrical contacting that quickly complicates 3D scalability. In contrast, large-angle TDBG is far simpler – the decoupling of individual bilayers is non-angle specific, and the bilayers are intrinsically gapped by crystal fields. This makes it possible to build thicker photoconductors with alternating layers of hBN and TDBG (inset Fig. 3). Moreover, with recent improvements in CVD growth[28] and automated transfers[29], our device architecture has a clear potential for upscale and can be used to engineer a new class of 3D photoconductors with 2D material properties.

## Methods
### Device fabrication of TDBG transistor
The samples are fabricated using the 'cut-and-stack' method, similar to previously reported 'tear-and-stack' technique[30]. Typically, a thin hBN flake (~10-15 nm) is picked using hot-pick up technique[31,32] using a polypropylene carbonate (PC) film on a polydimethylsiloxane (PDMS) stamp at 90º C. This hBN flake is then later used to pick up a part of pre-cut Bernal BLG flake, mechanically exfoliated on $Si^{++}/SiO_2$ (285 nm) from highly oriented pyrolytic graphite, and pre-characterized using optical microscopy, and Raman spectroscopy[33]. Subsequently, the remaining BLG flake is rotated to a target angle (~15º) and then picked up by the hBN-BLG stack on the PC film. Finally, the stack is used to pick up a last layer of hBN and later dropped on a pre-patterned marker chip of $Si^{++}/SiO_2$ (285 nm) at 180ºC, squeezing out the bubbles, and impurities as previously reported. The stack is then shaped into a Hall bar geometry using $SF_6$ plasma, and $O_2$ plasma to etch top hBN, and BLG respectively[31], and further metalized using 3/15/30 nm of Cr/Pd/Au. For the devices with top gate, a top stack of hBN-graphite-hBN was prepared, patterned and contacted similar to the bottom stack.

### Low-temperature infrared photocurrent measurements
The MIR measurements were performed at the temperature of 4K, unless otherwise noted. The samples were placed in a Montana Instruments optical cryostat with AR-coated ZnSe windows. The infrared beam was emitted by a Daylight MIRcat tunable QCL laser. The wavelength used was 11.07 µm, unless otherwise noted. The beam was expanded to ~13 mm diameter by a set of parabolic mirrors. The power was controlled by rotating a ZnSe holographic polarizer and the polarization was kept unaltered by a third, fixed ZnSe holographic polarizer. The light was then focused on the sample with a 15 x reflective objective (NA = 0.5). The focus spot is typically 20 µm diameter for 11.07 µm excitation.

The position of the focal point was scanned by moving the objective with stepper motors, with sub-micron resolution. The resulting displacement of the objective's optical axis from the beam's optical axis is negligible on the scale of the beam diameter.

The photoconductivity was measured by taking the difference of the sample conductivity with and without illumination. For measuring the 2-point probe and 4-point probe resistivity we used two SR860 lock-ins simultaneously. The sample was biased with 500 µV AC at 13.1 Hz by lock-in and the gate voltage was applied with a Keithley 2400 Sourcemeter.

The DC measurements in Fig. 4c were performed by biasing the sample and reading the current with another Keithley 2400 Sourcemeter. The 2-point probe voltage was measured by a DAQ after amplification and filtering by ITHACO 1201 low noise voltage preamplifier.

### Low-temperature broadband photocurrent measurements
The broadband photocurrent measurements presented in Fig. 2a were performed in a modified FTIR setup. A liquid 4He cooled optical cryostat was coupled with a commercial Bruker FTIR, where the laser was redirected on the TDBG detector instead of FTIR's internal DTGS detector. For the IR (2 to 20 µm) measurement, FTIR's internal thermal light source Globar was shined instead of QCL. In the case of the THz detection, a cold low-pass THz filter was used to ensure the low temperature of the TDBG detector.

### Data availability
The data that support the plots within this paper and other findings of this study are available from the corresponding authors upon reasonable request.

**Acknowledgements**

We thank D. B. Ruiz, S. Castilla, D. De Fazio, M. Amir Ali, G. Li, and I. Torre for technical discussions. We further thank M. Ceccanti for making the illustration presented in Fig. 1a. H.A., K. N., and R. B. acknowledges funding from the European Union's Horizon 2020 research and innovation program under the Marie Skłodowska-Curie grant agreement No. 665884, 713729, and 847517, respectively. S. B. P. acknowledges funding from the "Presidencia de la Agencia Estatal de Investigación" within the PRE2020-094404 predoctoral fellowship. G.S. and A.F. gratefully acknowledges funding from the ERC Grant CHIC (No. 724344) and J. Faist for discussions. F.H.L.K. acknowledges support from the ERC TOPONANOP (726001), the government of Spain (PID2019-106875GB-I00; Severo Ochoa CEX2019-000910-S [MCIN/AEI/10.13039/501100011033], PCI2021-122020-2A funded by MCIN/AEI/10.13039/501100011033) and by the "European Union NextGenerationEU/PRTRFundació Cellex, Fundació Mir-Puig, and Generalitat de Catalunya (CERCA, AGAUR, SGR 1656). Furthermore, the research leading to these results has received funding from the European Union's Horizon 2020 under grant agreement no. 881603 (Graphene flagship Core3) and 820378 (Quantum flagship). A.P. acknowledges support from the European Union's Horizon 2020 research and innovation programme under the Marie Sklodowska-Curie grant agreement No 873028 and from the Leverhulme Trust under the grant agreement RPG-2019-363. R.K.K. acknowledges the EU Horizon 2020 program under the MarieSkłodowska-Curie grants 754510 and 893030 and the FLAG-ERA grant (PhotoTBG), by ICFO, RWTH Aachen and ETHZ/Department of Physics. A. B acknowledges support from ERC Advanced Grant No. 692876, MICINN Grant No. RTI2018-097953-B-I00 and PID2021-122813OB-I00, AGAUR (Grant No. 2017SGR1664), the Fondo Europeo de Desarrollo, the Spanish Ministry of Economy and Competitiveness through Quantum CCAA, EUR2022-134050, and CEX2019-000910-S [MCIN/AEI/10.13039/501100011033], MCIN with funding from European Union NextGenerationEU(PRTR-C17.I1), Fundacio Cellex, Fundacio Mir-Puig, Generalitat de Catalunya through CERCA. K.W. and T.T. acknowledge support from the Elemental Strategy Initiative conducted by the MEXT JPMXP0112101001) and JSPS KAKENHI JP19H05790 and No. JP20H00354).

## Supplementary Section 1 – Twisted Double Bilayer Device Characterization

In large angle twisted double bilayer graphene, individual layers are electronically de-coupled from one another due to the large mismatch in momentum space. As a result, each bilayer experiences a strong crystal field effect because they lie in an asymmetric dielectric environment, with hexagonal boron-nitride and a separate electronically de-coupled bilayer graphene encapsulating either side of the material. This asymmetric environment result in a crystal field that inflicts a displacement field in the bilayer graphene, breaking inversion symmetry and gapping out its spectrum. The crystal field is intrinsic to the heterostructure and does not require any further electrostatic potentials to be applied as in the case of stand-alone bilayer graphene. To demonstrate the robustness of the crystal field gap, Figure. S1a plots the resistivity as a function of gate voltage measured in another one of our dual-gated TDBG devices. It shows similar behaviour to the single-gated device presented in the main text Fig. 1c, including a sharp resistivity peak at the charge neutrality point that shows activation behaviour upon increasing temperature, revealing a gap of 8 meV.

To further characterize our devices, we performed magneto transport measurements in dual-gated devices that allow us to control the carrier density in each bilayer. Because the layers should be de-coupled, a dual-gating scheme allows us to dope one drastically different to the other so that the transport properties of each layer can be dis-entangled. Figure. S1b plots the top gate vs bottom gate map of the resistivity measured in one of our dual-gated devices. Notably, there exists two high resistance lines that mark the zero density regimes in each of the two bilayers, demonstrating their electronic decoupling. The decoupling of crystal layers is further evident in the magnetoresistance. Fig. S1c plots maps of the resistivity as a function of magnetic field $B$ and carrier density $n$; the gate voltages are swept together along carrier densities with constant displacement fields. In contrast to stand-alone bilayer graphene, we find two sets of Landau levels emerge at different gate voltages. Each set corresponds to the neutrality points in each bilayer, further demonstrating the electronic decoupling of the crystal layers.

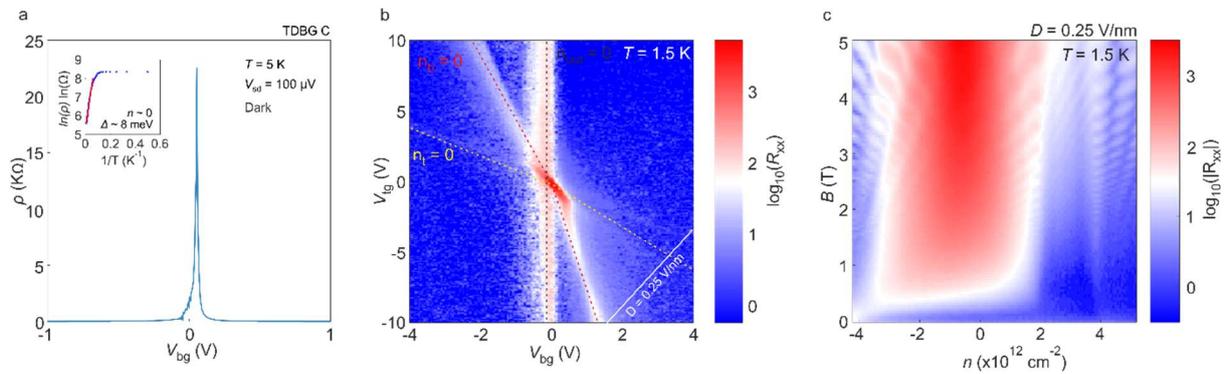

**Figure S1: TDBG transport characterization. a,** Dark resistivity (ρ) as a function of back-gate ($V_{bg}$). Inset top left: Temperature (T) dependence of ρ measured at the CNP showing activation behaviour. **b,** Logarithmic four-probe resistance as a function of both $V_{tg}$ and $V_{bg}$, with a solid white line indicating D = +0.25 V/nm. **c,** Logarithmic four-probe resistance as a function of charge carrier density and out-of-plane magnetic field, measured for a fixed D = +0.25 V/nm.

## Supplementary Section 2 – Giant IR photoresponse in other TDBG devices

Figure. S2 plots IR photoconductivity measurements performed in another one of our TDBG devices. In the dark, the device exhibits a strong peak in the resistivity due to the gapped state. Upon photoexcitation, the resistivity drops dramatically (20 kΩ) indicating a strong photoconductive effect.

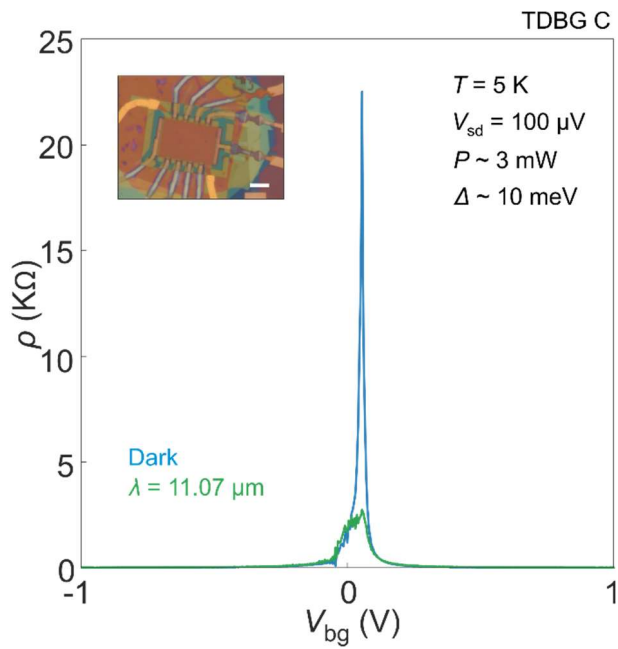

**Figure S2: Photoconductivity in another TDBG device.** ρ as a function of $V_{bg}$ measured in dark and under illumination. Inset top left: Optical picture of the TDBG device with scale bar of 10 μm.

## Supplementary Section 3– Absorption Calculations and Detector parameter estimates

### Absorption Calculation at 11.07 μm wavelength excitation

To calculate the internal quantum efficiency in the main text, we first must calculate the absorption coefficient of our TDBG heterostructure that includes encapsulating hexagonal-boron nitride layers and graphite gates (Fig. S3a). For this, we used the Rigorous Coupled Wave Analysis (RCWA). This method considers the thicknesses (obtained with AFM measurements) of each material in the layered system and solves Maxwell's equations with incoming plane waves. Inside each layer, we solve the electromagnetic problem with a semi-analytical technique: we pick the growth direction analytically while expanding the in-plane fields using a discrete Fourier transform. Once the wave propagation inside each layer is solved, we connect the solutions according to boundary conditions using the scattering matrix of each layer. With this technique, we can fully compute our devices' optical response, extracting both the scattering parameters (reflection, transmission, absorption) and the electric field distribution inside the structure.

The optical conductivity of each layer was determined in the IR range corresponding to wavelength excitations 2-12 μm. For hBN we used the mid-IR dielectric function of hBN as reported in ref (1). For graphite, we derived the dielectric function as follows. We assume that the volume current density flowing in graphite can be approximated as the current flowing in adjacent uncoupled layers of graphene. To do that, we substitute the surface current density in graphene $j_{\parallel}(\omega) =$

$\sigma_{graphene}(\omega).E_\parallel(\omega)$ with a volume current density $I_\parallel(\omega) = j_\parallel(\omega)/\delta$ where $\sigma_{graphene}(\omega)$ is the local RPA optical conductivity obtained by the Kubo Formula[2] , $E_\parallel(\omega)$ is the in-plane component of the electric field, and δ is the thickness of the graphite layer. In this way, the three-dimensional conductivity for graphite becomes $\sigma_{graphite}(\omega) = \sigma_{graphene}(\omega)/\delta$ and its electromagnetic behaviour can be described by its dielectric function $\epsilon_{graphite}(\omega)$, which is connected to the conductivity via the general relation[3]

$$\varepsilon(\omega) = 1 + \frac{i\sigma(\omega)}{\delta\varepsilon_0(\omega)} \tag{S1}$$

In large-angle TDBG, the two BLG layers are decoupled (See Supplementary Section 2) . Therefore, to account for its optical response in the simulations, we consider two identical sheets of BLG, separated by a thin layer of vacuum with thickness = 0.34 nm[4,5]. The behaviour of every single BLG is calculated with the help of equation (S1), with thickness = 0.68 nm and the optical conductivity for $\sigma_{BLG}(\omega)$ is obtained from band structure calculations. The substrate is constituted by 285 nm of Silicon dioxide (SiO$_2$)/p-doped Si, with dielectric function for SiO$_2$ accounting for the optical phonons from E. D. Palik[6] and permittivity for p-doped Si from J. W. Cleary et al.[7] From the scattering parameters extracted by the RCWA simulation, we can reconstruct the absorption spectrum of our device. Figure S3b shows the percentage of light absorption for different frequencies in the active area of the system, calculated as the sum of the absorptions in each of the three layers used to model the TDBG (see the red highlighted box in Fig. S3a). We calculate α = 0.944 % at the laser frequency of 11.07 µm = 903.34 cm$^{-1}$.

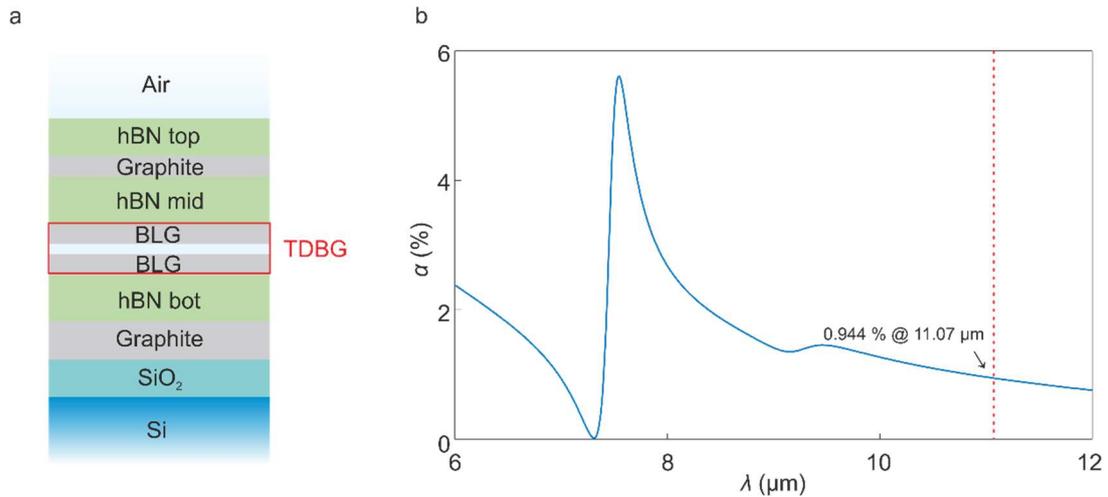

**Figure S3: Absorption calculation. a,** Schematics of the multi-layered structure used in the simulation of the TDBG device. The active photodetection area is highlighted in red and simulated as a three-layer system made of BLG-vacuum-BLG. The thicknesses of the layers are not to scale. **b,** Optical absorption spectrum for the TDBG obtained with RCWA simulations. The red line indicates the operation wavelength (λ = 11.07 µm) of our experiment in main text Fig. 3.

**Noise Equivalent power (NEP)**

In our measurements the NEP was set-up limited rather than device limited (see Fig. 1e of main text). The device limited NEP is limited mostly by shot noise (Johnson noise is an order of magnitude smaller in our system at 5 K).

$$Total\ noise = \sqrt{{S_{sho}}^2 + {S_{johnson}}^2}$$

where $S_{sho} = 2eI_{dark}$ = 2.53E-13. By using the linear power dependence in Fig. 1e of main text, we translated total noise to an intrinsic device limited NEP of 66 fW/$\sqrt{Hz}$.

**Supplementary Section 4 – Terahertz photoconductivity in TDBG**

The spectral measurements in the terahertz regime (Fig. 2c of the main text) revealed a broadband terahertz photoconductive effect in our TDBG heterostructures. Here, we present terahertz photoconductivity measurements obtained in one of our TDBG heterostructures using a single frequency excitation (2.5 THz) from a terahertz gas laser. Figure. S4 plots the voltage $\Delta V$ as a function of gate voltage $V_{bg} - V_{cnp}$ (V) measured in one of our devices. The measurement was performed in a double modulation like scheme in which a light modulation chopped at 337 Hz was mixed with an AC bias modulation at 19 Hz and the resulting signal read out at the difference frequency corresponded to the photoconductivity. The data shows a peaked response at the charge neutrality point (CNP) as well as some shoulder like feature, likely due to some charge inhomogeneity in the system. The inset plots a power dependence of the photoresponse performed at the CNP. The power corresponds to illuminated power on the device, and is much larger than the infrared measurements because of the large beam size and lower power density.

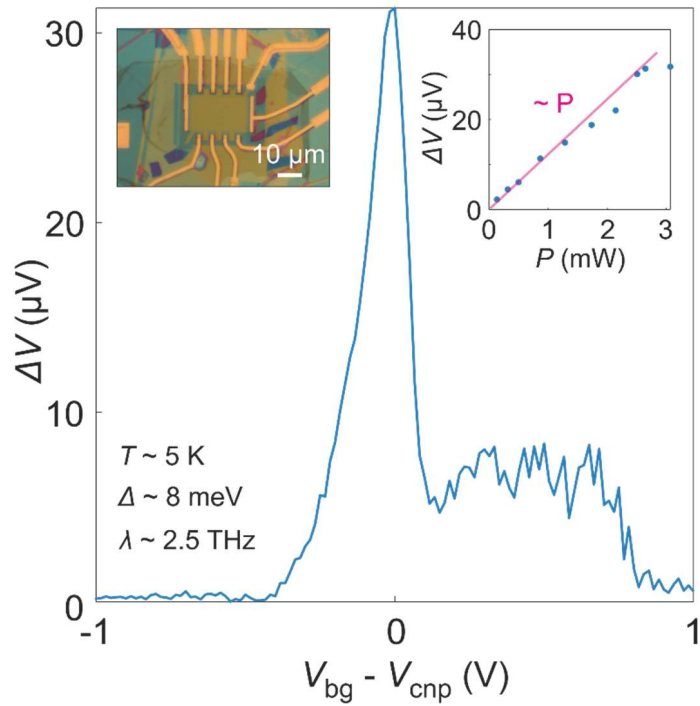

**Figure S4: THz photoconductivity in dual gated TDBG detector.** Two probe light induced photovoltage as a function of back-gate voltage measured with 20 µA constant current at 19 Hz AC modulation and 337 Hz chopper frequency. The left inset shows the device's optical picture with a scale bar of 10 μm. The right inset shows a linear power dependence measured for the same device as a function of illuminated power.

**Supplementary Section 5 – Transport characterization of dual gated bilayer graphene device**

To understand the large photoconductive response in our TDBG heterostructures, we studied the photoconductivity in dual-gated bilayer graphene heterostructures as a control experiment. The devices are encapsulated with hexagonal boron-nitride and gated with graphite gates to ensure the highest electronic quality (see optical image in inset of Fig. S5b). Figure. S5a plots a colour map of the four probe resistance $R_{xx}$ measured as a function of the top gate and bottom gate ($V_{tg}$ and $V_{tg}$ resistively). It shows the usual behaviour for bilayer graphene where the resistivity at the charge neutrality point increases due to a gap opening induced by the displacement field that breaks inversion symmetry[8,9]. Figure S5b plots temperature dependence of the resistance showing activation behaviour from which extract the gap values or around 13 meV).

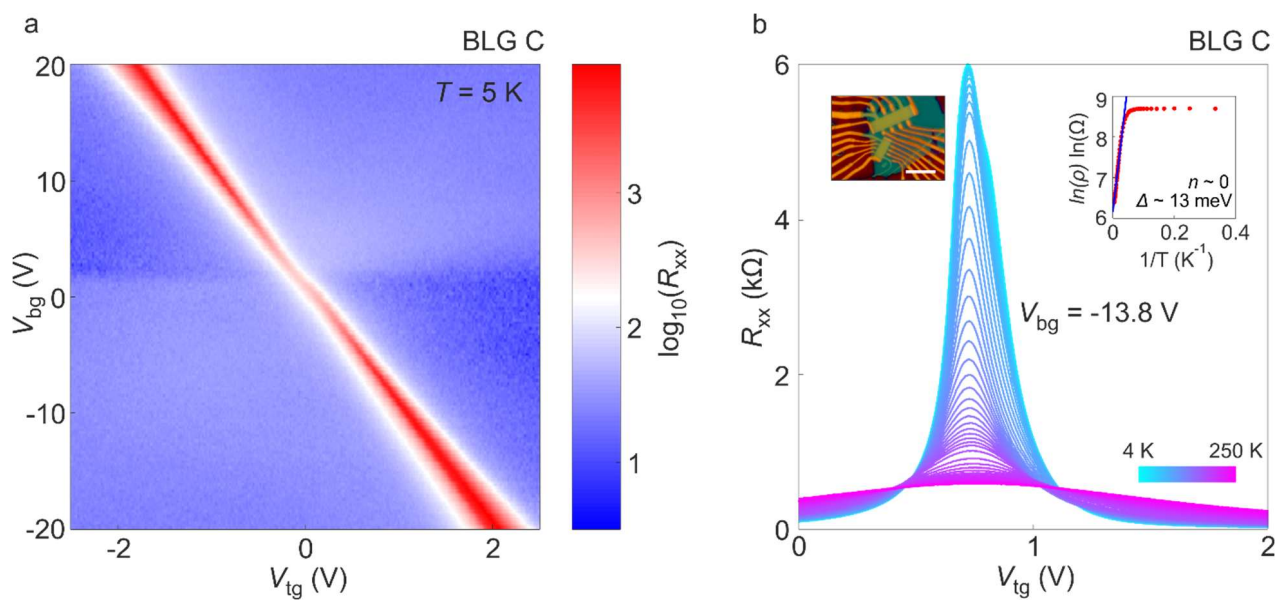

**Figure S5: Transport measurement of one of the Dual gated bilayer graphene device. a,** Logarithmic four-probe resistance ($R_{xx}$) as a function of top gate ($V_{tg}$) and back gate ($V_{bg}$). **b,** Four-probe resistance as a function of the top gate $R_{xx}(V_{tg})$ for a fixed $V_{bg}$ of -13.8 V, measured at different temperatures. Top inset left: Optical picture of two dual gated bilayer graphene devices, with a scale bar of 10 μm. Top inset right: Temperature dependence of four-probe resistance at CNP showing activation behaviour.

**Supplementary Section 6 – Photoconductor Vs Bolometric mechanism – dis-entangling through measurements of IV characteristics**

As described in the main text, the enhancement in conductivity via photoexcitation can be attributed to two mechanisms, either the photoconductive or bolometric response. Here we present a qualitative model that shows how the two mechanisms lead to different IV characteristics in the photoresponse of gapped systems.

We present a toy model of the electron transport in our gapped device by considering two contributions including the ballistic transmission of charge carriers across the gap, and a diffusive contribution described by free charge carriers that lie in states above the gap.

$$J = \int_{\varepsilon_F - eV}^{\epsilon_F} T(\epsilon, V_{Bias}) d\epsilon + n_e \mu_e e \frac{V_{Bias}}{d} + n_h \mu_h e \frac{V_{Bias}}{d}$$

The first term corresponds to the ballistic transmission with $T(\epsilon, V)$ representing the transmission coefficients that depends on the Fermi energy ($\epsilon$) and bias voltage (*V*). The second and third terms correspond to the diffusive Drude current where $n_{e/h}$ the electron/hole carrier density, $\mu_{e/h}$ the electron/hole mobility and $V_{Bias}$ the voltage applied to the sample.

The difference between photoconductivity and bolometric lies in the way in which the free carrier density $n_{e/h}$ is generated in the sample upon photoexcitation. For a bolometer, $n_e$ increases through temperature smearing of the Fermi-Dirac distribution.

$$n_{e/h} = DOS \frac{e}{\hbar} f(\varepsilon) d\varepsilon$$

$$f(\varepsilon) = \frac{1}{1 + e^{\frac{(\varepsilon - \epsilon_F)}{k_b T}}}$$

Consequently, for small temperature changes (low powers) the main effect is the lowering of threshold voltage for ballistic transmission, whereas the current around zero bias remains little affected (see Fig. 4a). In contrast, for a photoconductor, the free carrier density is governed by the absorption of photons and is proportional to the power (*P*)[10].

$$n_{e/h} = \alpha \frac{P}{S_{laser}} \frac{M\tau}{\hbar \omega}$$

Here, $\alpha$ is light absorption, *S*$_{laser}$ the area of laser spot, *M* is a carrier multiplication factor, $\tau$ is a phenomenological carrier lifetime, and $\hbar\omega$ is photon energy.

This allows for a dominating contribution from the diffusive current even for small excitation powers, and the current around zero bias increases linearly with power. We note that, even in the case of a finite dark current (due to thermal generation), the differences still prevail except for a finite conductivity off-set in the differential resistance measurements.

**Supplementary Section 7 – Comparison of photoconductivity with Silicon gated and graphite gated devices**

As demonstrated in figure 5 of the main text, electron-hole collisions are strongly suppressed in our TDBG heterostructures compared to stand-alone bilayer graphene. As was shown recently, local metallic gates in graphene-based heterostructures have been shown to strongly screen electron-electron interactions and might explain the origin of supressed electron-hole interactions in our devices. However, an estimate of the Thomas-Fermi screening length in bilayer graphene shows that gates must be placed only a few atomic layers away from active channel, in contrast to our TDBG devices which have graphite gates separated from the channel with 20 nm dielectric layers. Nonetheless, to verify that the screening does not originate from local graphite gates, we compared photoconductivity in bilayer graphene devices that were gated with global silicon back gates (where proximity effects are certainly minimal) and local graphite gates. Fig. S6a plots the dark conductivity $\sigma_{xx}$ as a function of carrier density *n* measured in the two devices with displacement fields applied to gap the systems to 10 meV. The main difference in electronic quality can be seen around zero doping, in which the gapped state is more well defined for graphite gated samples (lower conductivity) due to the cleaner electrostatic gating. Fig. S6b plots $\Delta\sigma_{xxs}$ (P) measured in two BLG devices of the same gap size (10 meV); the red data points plot measurements in a Si gated devices whereas the green plots measurements in a graphite gated device. Despite the improvements in electronic quality of the gapped state, the photoconductivity measurements show no marked difference between the two with only around a 10-20% variation in the photoconductivity. This tells us that the photoconductivity enhancements observed in our experiments do not come from the presence of local proximity screening gates but rather the intrinsic interlayer screening of the TDBG heterostructure.

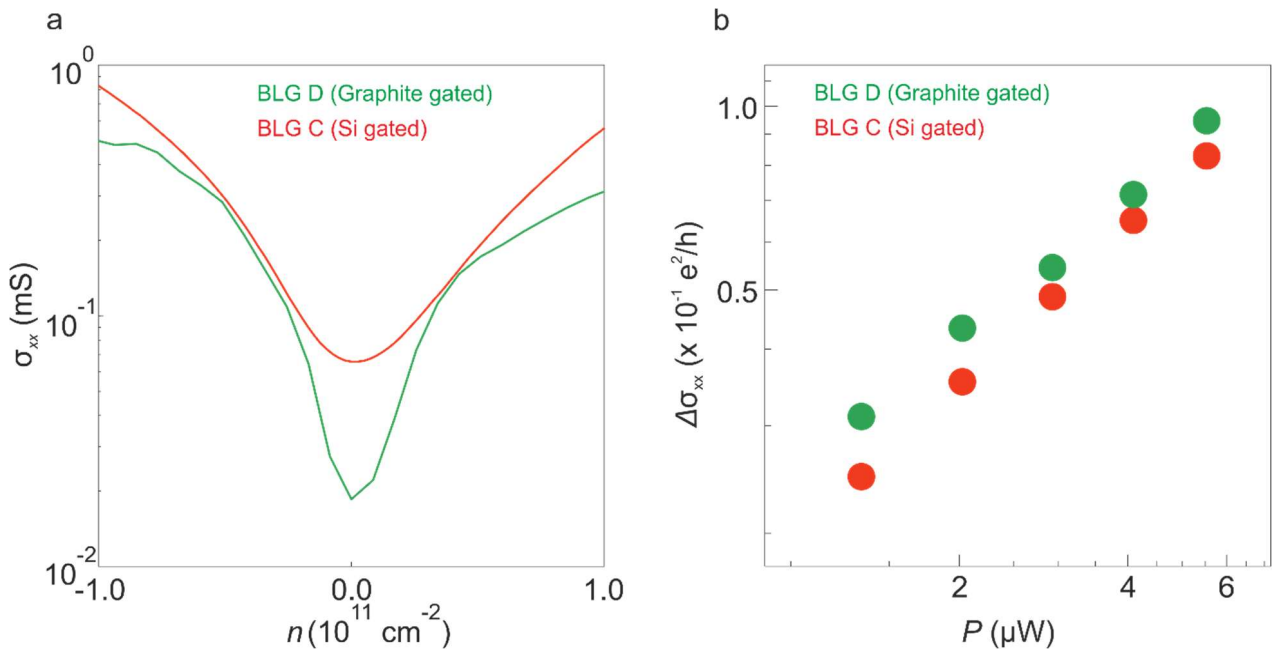

**Figure S6: Comparison of BLG devices with and without local graphite gate. a,** Dark conductivity $\sigma_{xx}$ as a function of carrier density *n* measured in the two BLG devices with displacement fields applied to gap the systems around 10 meV. The device with local graphite gate have significantly higher quality and lower $\sigma_{xx}$ around zero doping compared to the device with silicon back gate. **b,** Photoconductivity as a function of shined power for the two similar gapped BLG devices as shown in panel a. The measurement was done by following the same procedure as main text Fig. 3. Both devices showed similar photoconductivity values and power dependence in the linear power regime, ruling out the role of substrate in photoconductivity enhancement.

**Supplementary Section 8 – Estimate of Thomas-Fermi screening length and suppression factor of electron-electron scattering length.**

To understand whether electron-hole collisions could be suppressed in our TDBG heterostructure, we analyse the analytical formula derived in ref (11) that describes how the electron-electron scattering length is enhanced by placing a proximity-screening gate nearby.

$$l_{ee} \approx \frac{4\hbar v_F E_F}{\pi} \frac{1}{(k_B T)^2 \ln\left(\frac{2E_F}{k_B T}\right)} \left(\frac{1 + 2dq_{TF}}{2dq_{TF}}\right)^2$$

where $E_F$ refers to the Fermi energy, $v_F$ the Fermi velocity, $d$ the dielectric spacing between the channel and the gate, and $q_{TF}$ the Thomas-Fermi screening length. The term on the right hand side of the equation in brackets describes the enhancement factor caused by the proximity-screening gate and should be the same for electron-hole collisions it is also governed by long-range Coulomb interactions. Estimating the Thomas-Fermi screening length $q_{TF} = \frac{2e^2 m}{\hbar^2 \epsilon} \approx 0.32$, taking an upper bound for the effective mass ($m$) in bilayer graphene[12] of $0.024 m_e$ and a dielectric constant for $\epsilon$ of the vacuum = 1, we calculate the enhancement factor $(1 + 2dq_{TF})/2dq_{TF})^2 \sim 1.5 - 2$, depending on the choice of dielectric constant or effective mass. Note, that encapsulating hBN layers may also influence the dielectric screening. This analysis is only to demonstrate the enhancement considering screening via a proximity gate is possible in our experiments. For the full calculation, see (Supplementary Section 9).

**Supplementary Section 9 – Calculation of electron-hole collision limited conductivity and effects of interlayer screening**

We consider a bilayer graphene sheet in the presence of a displacement field, which opens a gap in the band structure. For simplicity, we use a two-band approximation and we take the energy dispersion to be given by $\epsilon_{k,\lambda} = \hbar^2 k^2 / 2m$ and $m$ is the electron effective mass (equal to 0.03 times the bare mass).

To calculate the resistivity at high temperature, due to electron-hole scattering, we consider the Boltzmann equation for the electron distribution function $f_{k,\lambda}(r,t)$, i.e

$$\partial_t f_{k,\lambda}(r,t) + v_{k,\lambda} \cdot \nabla_r f_{k,\lambda}(r,t) - eE \cdot \nabla_k f_{k,\lambda}(r,t) = \mathfrak{I}_{ee}[f_{k,\lambda}]$$
(1)

Where $E$ is the applied electric field, -e is the electron charge, $v_{k,\lambda} = \hbar^{-1} \nabla_k \epsilon_{k,\lambda}$ is the particle velocity, while $\mathfrak{I}_{ee}[f_{k,\lambda}(r,t)]$ is the collision integral of electron-electron interactions. $\mathfrak{I}_{ee}[f_{k,\lambda}(r,t)]$ Conserves both the total momentum and energy at each scattering event.

$$\mathfrak{I}_{ee}[f_{k_1,\lambda_1}] = \frac{1}{A^3} \sum_{k_2 \lambda_2} \sum_{\substack{k_3 \lambda_3 \\ k_4 \lambda_4}} W_{ee}(k_1 \lambda_1; k_2 \lambda_2; k_3 \lambda_3; k_4 \lambda_4) \delta(k_1 + k_2 - k_3 - k_4) \delta(\varepsilon_{k_1 \lambda_1} + \varepsilon_{k_2 \lambda_2} - \varepsilon_{k_3 \lambda_3} - \varepsilon_{k_4 \lambda_4}) \times [f_{k_1,\lambda_1} f_{k_2,\lambda_2}(1 - f_{k_3,\lambda_3})(1 - f_{k_4,\lambda_4}) - (1 - f_{k_1,\lambda_1})(1 - f_{k_2,\lambda_2}) f_{k_3,\lambda_3} f_{k_4,\lambda_4}]$$ (2)

Here,

$$W_{ee}(k_1 \lambda_1; k_2 \lambda_2; k_3 \lambda_3; k_4 \lambda_4) = \frac{2\pi}{\hbar} [V_{ee}(k_1 - k_3, \varepsilon_{k_1 \lambda_1} - \varepsilon_{k_3 \lambda_3})]^2 \mathcal{D}_{k_1 \lambda_1; k_3 \lambda_3} \mathcal{D}_{k_2 \lambda_2; k_4 \lambda_4},$$
(3)

Where $V(q, \omega) = v_q / \epsilon(q, \omega)$, $v_q = 2\pi e^2 / (\bar{\epsilon} q)$ is the bare Coulomb interaction, $\bar{\epsilon}$ is the dielectric constant of the surrounding environment, while $\epsilon(q, \omega)$ is the dielectric function of bilayer graphene in the two-band approximation (*i.e* the screening due to the free electrons themselves). In eq. 3, $\mathcal{D}_{k,\lambda;k`\lambda`}$ is the matrix element of the density operator between the states labelled by $k$ and $\lambda$, and $k'$ and $\lambda'$.

In the steady-state and to linear order in the electric field, we consider the following *ansatz* for $f_{k,\lambda}$:

$$f_{k,\lambda} = f_{k,\lambda}^{(0)} + \tau e v_k \lambda \cdot E \frac{\partial f_{k,\lambda}^{(0)}}{\partial \varepsilon_{k,\lambda}},$$

(4)

Where our goal is to determine $\tau$, and

$$f_{k,\lambda}^{(0)} = \frac{1}{e^{[\varepsilon_{k,\lambda} - \mu_\lambda]/(k_B T)} + 1},$$

(5)

Is the equilibrium distribution function. Note that the chemical potentials of electrons and holes, $\mu_\pm$, may not coincide in a general non-equilibrium situation (e.g., under external pumping). Eq. (1) thus becomes,

$$e v_{k,\lambda} \cdot E \left( -\frac{\partial f_{k,\lambda}^{(0)}}{\partial \varepsilon_{k,\lambda}} \right) = \mathfrak{I}_{ee}^{(lin)} [f_{k,\lambda}(r,t)],$$

(6)

where $\mathfrak{I}_{ee}^{(lin)} [f_{k,\lambda}(r,t)]$ is the linearized electron-electron collision integrals. We multiply Eq. (6) by $v_{k,\lambda}$ and sum over $k$ and $\lambda$. From the left-hand side we get $\mathcal{D}eE$, where

$$D = \frac{1}{2A} \sum_{k,\lambda} \left( -\frac{\partial f_{k,\lambda}^{(0)}}{\partial \epsilon_{k,\lambda}} \right) |v_{k,\lambda}|^2 = \frac{N_F}{\pi} \frac{k_B T}{\hbar^2} \mathcal{F}_D(\bar{\Delta}),$$

(7)

Where $\mathcal{F}_D(\bar{\Delta}) \equiv \ln[2 \cosh(\bar{\Delta})] - \bar{\Delta}\tanh(\bar{\Delta})$ and $\bar{\Delta} = \Delta/(k_B T)$. In the limit of large temperature ($\bar{\Delta} \to 0$), $\mathcal{F}_D(\bar{\Delta}) \to \ln(2)$. From the right-hand side, we get $I_{ee} \tau e E$, where

$$I_{ee} = \frac{2}{8 \hbar k_B T \mathcal{A}^4} \sum_{\substack{k_1 \lambda_1 \\ k_2 \lambda_2}} \sum_{\substack{k_3 \lambda_3 \\ k_4 \lambda_4}} |V_{ee}(k_1 - k_3, \varepsilon_{k_1,\lambda_1} - \varepsilon_{k_3,\lambda_3})|^2 \mathcal{D}_{k_1 \lambda_1; k_3 \lambda_3} \mathcal{D}_{k_2 \lambda_2; k_4 \lambda_4} \delta(k_1 + k_2 - k_3 -$$
$$k_4) \times \delta(\varepsilon_{k_1,\lambda_1} + \varepsilon_{k_2,\lambda_2} - \varepsilon_{k_3,\lambda_3} - \varepsilon_{k_4,\lambda_4}) f_{k_1,\lambda_1}^{(0)} f_{k_2,\lambda_2}^{(0)} \left(1 - f_{k_3,\lambda_3}^{(0)}\right) \left(1 - f_{k_4,\lambda_4}^{(0)}\right) |v_{k_1,\lambda_1} + v_{k_2,\lambda_2} - v_{k_3,\lambda_3} - v_{k_4,\lambda_4}|^2.$$
(8)

The transport time then becomes $\tau^{-1} = D^{-1} I_{ee}$. The conductivity is obtained by calculating the current $j = -e \sum_{k,\lambda} v_{k,\lambda} f_{k,\lambda}$ and equating it to $\sigma E$. We obtain

$$\sigma = \frac{e^2 D^2}{I_{ee}}$$

(9)

We now consider the case in which electrons and holes scatter within the same band, thus $\lambda' = \lambda$ and $\eta' = \eta$ (but $\eta \neq \lambda$). We can then approximate the matrix elements of the density operator with $D_{k,\lambda;k',\lambda'} \cong 1$. Furthermore, we approximate the Coloumb interaction with the statically screened one,

$$V_{ee}(q,\omega) = V_{ee}(q) = \frac{2\pi e^2}{\bar{\epsilon}(q+q_{TF})},$$

(10)

Where

$$q_{TF} = \frac{2\pi e^2}{\bar{\epsilon}} \frac{1}{\mathcal{A}} \sum_{k,\lambda} \left(-\frac{\partial f_{k,\lambda}^{(0)}}{\partial \epsilon_{k,\lambda}}\right) = \frac{2\pi e^2}{\bar{\epsilon}} \frac{mN_F}{2\pi\hbar^2} \mathcal{F}_{TF}(\bar{\Delta}),$$

(11)

Is the Thomas-Fermi screening wave vector. Here,

$$\mathcal{F}_{TF}(\bar{\Delta}) = \int_0^\infty \frac{dx}{\cosh^2(\sqrt{\Delta^2+x^2})}$$

(12)

In the limit of large temperatures, $\mathcal{F}_{TF}(\bar{\Delta} \to 0) \to 1$. We find that, in general, $q_{TF}$ is always larger than the typical values taken by $q$. Thus, we will neglect $q$ in Eq. (10). Using that $f_{k_1,\lambda_1}^{(0)}\left(1 - f_{k',\lambda'}^{(0)}\right) = -n^{(0)}(\varepsilon_{k,\lambda} - \varepsilon_{k',\lambda'})(f_{k_1,\lambda_1}^{(0)} - f_{k',\lambda'}^{(0)})$ where $n^{(0)}(\omega) = [e^{\omega/(k_BT)} - 1]^{-1}$ is the Bose distribution, and introducing $q = k_1 - k_3$ and $\omega = \varepsilon_{k_1,\lambda_1} - \varepsilon_{k_3,\lambda_3}$, we rewrite Eq. (8) as

$$I_{ee} = \frac{\pi}{16\hbar k_B T \mathcal{A}^3} \left(\frac{2\pi e^2}{\bar{\epsilon} q_{TF}}\right)^2 \sum_q \int_{-\infty}^\infty d\omega \frac{1}{\sinh^2\left(\frac{\omega}{2k_BT}\right)} \sum_{k,k'} \partial(\varepsilon_{k,+} - \varepsilon_{k-q,+} - \omega)\partial(\varepsilon_{k',-} - \varepsilon_{k'+q,-} + \omega) \times$$

$$(f_{k,+}^{(0)} - f_{k-q,+}^{(0)})(f_{k',-}^{(0)} - f_{k'+q,-}^{(0)}) \left|v_{k,+} + v_{k',-} - v_{k-q,+} - v_{k'+q,-}\right|^2$$

(13)

At low temperatures, the integrand in Eq. (13) is peaked at $\omega = 0$. In this case, we can approximate

$$\left|v_{k,+} + v_{k',-} - v_{k-q,+} - v_{k'+q,-}\right|^2 = \frac{4\hbar^2 q^2}{m^2}$$

(14)

After some lengthy but straightforward manipulations, we can rewrite

$$I_{ee} = -\hbar \left(\frac{N_F}{\pi} \frac{k_B T}{\hbar^2} \mathcal{F}_D(\bar{\Delta})\right)^2 \frac{2\pi^2}{3N_f^2} \frac{\mathcal{F}_1(\bar{\Delta})}{\mathcal{F}_{TF}^2(\bar{\Delta})\mathcal{F}_D^2(\bar{\Delta})}$$

(15)

Where

$$\mathcal{F}_1(\bar{\Delta}) = \int_0^\infty dx \frac{1}{\cosh^2(\sqrt{\Delta^2+x^2})} \int_x^\infty dx' \frac{1}{\cosh^2(\sqrt{\Delta^2+x'^2})} \text{arcsinh}\left(\frac{x}{\sqrt{x'^2-x^2}}\right)$$

(16)

In the limit of large temperatures, $\mathcal{F}_1(\bar{\Delta} \to 0) \to 0.28$. Thus, the conductivity, obtained by plugging the results in Eq. (7) and (15) into Eq. (9), is

$$\sigma = \frac{e^2}{h} \frac{3N_f^2}{\pi} \frac{\mathcal{F}_{TF}^2(\bar{\Delta})\mathcal{F}_D^2(\bar{\Delta})}{\mathcal{F}_1(\bar{\Delta})}$$

Figure S7 shows a plot of the function $\mathcal{F}_\sigma(\bar{\Delta}) \equiv \mathcal{F}^2_{TF}(\bar{\Delta})\,\mathcal{F}^2_D(\bar{\Delta})/\mathcal{F}_1(\bar{\Delta})$ as a function of $\Delta^{-1} = k_B T/\Delta$. At large temperatures, $\mathcal{F}_\sigma(\bar{\Delta} \to 0) \to 1.72$, thus, for an isolated bilayer graphene, we get that $\sigma \to 26\, e^2/h$. This value can of course vary slightly depending on the parameters used. Nonetheless, when two bilayers are placed in close proximity, the number of fermion flavours doubles, i.e $N_f \to 2N_f$. In turn, both the density of states, (Eq.7) and the Thomas-fermi wave vector, Eq. (11), double, while $I_{ee}$ remains unchanged. The latter result stems from the cancellation of $N_f$ between the number of scattering channels, which increases and would quadruple when $N_f$ doubles, the net effect is that $I_{ee}$ scales as $q_{TF}^{-1}$ [see, e.g., Eq. (13)] and is therefore reduced by a factor 4 when $N_f$ doubles. The net effect is that $I_{ee}$ does not change with the addition of the second bilayer. As a result, the conductivity in Eq. (17) quadruples.

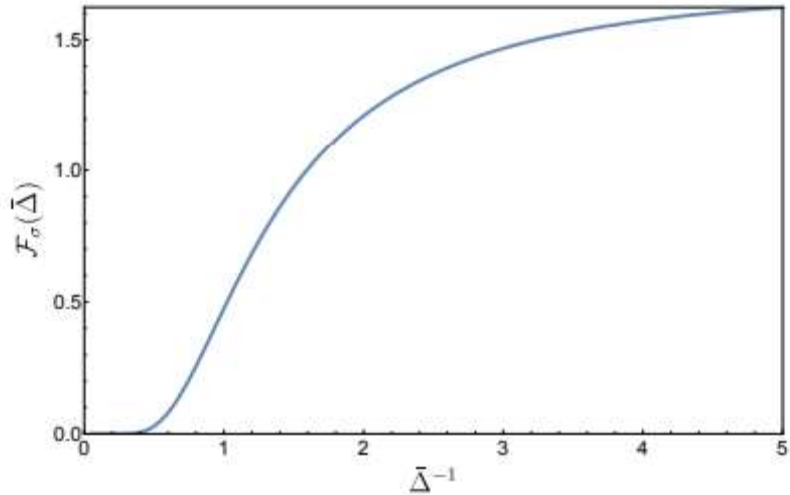

**Figure. S7:** The function $\mathcal{F}_\sigma(\bar{\Delta}) \equiv \mathcal{F}^2_{TF}(\bar{\Delta})\,\mathcal{F}^2_D(\bar{\Delta})/\mathcal{F}_1(\bar{\Delta})$ as a function of $\Delta^{-1} = k_B T/\Delta$.